\documentclass[twocolumn,aps,nofootinbib]{revtex4}
\usepackage{amsfonts, amssymb}
\usepackage{graphicx, epsfig,bm}
\newcommand{\be}{\begin{equation}}
\newcommand{\ee}{\end{equation}}
\newcommand{\bea}{\begin{eqnarray}}
\newcommand{\eea}{\end{eqnarray}}

\newcommand{\pp}{\,\, .}
\newcommand{\vv}{\,\, ,}

\newcommand{\ApJ}{{Astrophys. J.\,}}

\def\fun#1#2{\lower3.6pt\vbox{\baselineskip0pt\lineskip.9pt
\ialign{$\mathsurround=0pt#1\hfill##\hfil$\crcr#2\crcr\sim\crcr}}}

\newcommand\lsim{\mathrel{\rlap{\lower4pt\hbox{\hskip1pt$\sim$}}
    \raise1pt\hbox{$<$}}}
\newcommand\gsim{\mathrel{\rlap{\lower4pt\hbox{\hskip1pt$\sim$}}
    \raise1pt\hbox{$>$}}}

\def\dslash{\not{\hbox{\kern-2pt $\partial$}}}
\def\Dslash{\not{\hbox{\kern-4pt $D$}}}
\def\Oslash{\not{\hbox{\kern-4pt $O$}}}
\def\Qslash{\not{\hbox{\kern-4pt $Q$}}}
\def\pslash{\not{\hbox{\kern-2.3pt $p$}}}
\def\kslash{\not{\hbox{\kern-2.3pt $k$}}}
\def\qslash{\not{\hbox{\kern-2.3pt $q$}}}

 \newtoks\slashfraction
 \slashfraction={.13}
 \def\slash#1{\setbox0\hbox{$ #1 $}
 \setbox0\hbox to \the\slashfraction\wd0{\hss \box0}/\box0 }

\def\ee{\end{equation}}
\def\be{\begin{equation}}

\begin{document}
\setlength{\unitlength}{1mm}
\title{Cosmological bounds on dark-matter--neutrino interactions}

\author{Gianpiero Mangano$^1$, Alessandro Melchiorri$^2$, Paolo
Serra$^2$, Asantha Cooray$^3$, and Marc Kamionkowski$^4$}

\affiliation{$^1$Physics Department and Sezione INFN, University of
Naples ``Federico II'', Via Cintia, 80126 Naples, Italy}

\affiliation{$^2$Physics Department and Sezione INFN, University of
Rome ``La Sapienza'', P.le Aldo Moro 2, 00185 Rome, Italy}

\affiliation{$^3$Center for Cosmology, Department of Physics and Astronomy,
  4129 Frederick Reines Hall, University of California, Irvine,
CA 92697}

\affiliation{$^4$California Institute of Technology, Mail Code
     130-33, Pasadena, CA 91125}


\begin{abstract}
We investigate the cosmological effects of a neutrino
interaction with cold dark matter.  We postulate a neutrino that
interacts with a ``neutrino interacting dark matter'' (NIDM)
particle with an elastic-scattering cross section that either
decreases with temperature as $T^2$ or remains constant with
temperature. The neutrino--dark-matter
interaction results in a neutrino--dark-matter fluid with
pressure, and this pressure results in diffusion-damped
oscillations in the matter power spectrum, analogous to the
acoustic oscillations in the baryon-photon fluid.
We discuss the bounds from the Sloan Digital Sky
Survey on the NIDM opacity (ratio of cross section to
NIDM-particle mass) and compare with the constraint from
observation of neutrinos from supernova 1987A.  If only a
fraction of the dark matter interacts
with neutrinos, then NIDM oscillations may affect current
cosmological constraints from measurements of galaxy clustering.
We discuss how detection of NIDM oscillations would suggest a
particle-antiparticle asymmetry in the dark-matter sector.
\end{abstract}
\bigskip

\maketitle

\section{Introduction}

Flat cosmological models with baryons (about $5\%$ of the total
energy content of the universe), cold dark matter (CDM, $25\%$),
cosmological constant (or dark energy, $70\%$) and an adiabatic,
nearly scale-invariant spectrum of density fluctuations explain
most cosmological observations. However, we still lack a
satisfactory understanding of both dark matter and dark energy,
a puzzle for both particle physics and cosmology.

The most favored candidates for dark matter are cold,
collisionless massive particles, which are non-relativistic for
most of the history of the Universe and so can cluster
gravitationally during matter domination. Candidates for these
dark-matter particles can be found in supersymmetric extensions of
the standard electroweak model---namely neutralinos with mass on
the order of 100 GeV \cite{Jungman:1995df}---or in other theories
(e.g., the axion, which may arise in the Peccei-Quinn mechanism
\cite{axion}). These cold-dark-matter models account well for
cosmic microwave background (CMB) observations on the largest
scales as well as measurements of the large-scale distribution of
galaxies.

However, observations on galactic and sub-galactic scales may
conflict with the predictions, from numerical simulations and
analytic calculations, of CDM models.  Indeed, cold and
collisionless dark-matter models seem to predict an excess of
small-scale structures \cite{silk1}, and numerical simulations
\cite{prada} predict far more satellite galaxies in the Milky
Way halo than are observed.

Several solutions have been proposed to explain these
discrepancies, for example, inflationary models with broken scale
invariance \cite{marcandrew}.  However, most other explanations
invoke modifications to the properties of dark-matter particles.
For example, a warm-dark-matter candidate, like a sterile
neutrino, has been suggested because it suffers free streaming and
suppresses the matter power spectrum on small scales~\cite{silk2}.
A dark-matter particle that results from decay of a short-lived
charged particle can also suppress small-scale power \cite{kris}.
Other possibilities include a dark-matter particle that interacts
with others particles such as  photons \cite{bom1,xuelei,dipole}
or neutrinos \cite{bom2,stefano} or self-interacting dark matter
\cite{spergel}. For a review of different alternative scenarios to
standard collisionless cold dark matter, see Ref.~\cite{tesina}.

In this paper, we investigate the possibility of a neutrino
interacting dark matter (NIDM) component. If dark matter and
neutrinos interact, there was an epoch in the very early universe
in which they were strongly coupled. Dark-matter perturbations
that entered the horizon during this period would then be erased
because of diffusion damping, and the suppression scale will
depend on the dark-matter--neutrino interaction. Even
if only a fraction of the dark matter interacts with neutrinos, a
pattern of oscillations in the matter power spectrum arises, much
like the oscillations in the baryon-photon fluid.

In the following, we limit our study of DM-neutrino couplings to
effects on cosmological scales in the frequency range smaller than
$k < 0.2\,h$\,Mpc$^{-1}$---i.e., on scales where linear
perturbation theory is viable. We consider flat cosmological
models with an adiabatic and nearly scale invariant spectrum
$P(k)\sim k^n$ of density perturbations where $n=0.97$. Unless
explicitly stated, the energy content of the Universe corresponds
to the standard $\Lambda$CDM model with baryons contributing as
$\Omega_b=0.05$ and a cold-dark-matter energy density
$\Omega_{dm}=0.25$. We also choose a Hubble parameter $h=0.73$, a
standard value $3.04$ for the effective number of (massless)
neutrinos \cite{Mangano:2005cc}, and dark energy in the form of a
cosmological constant with equation-of-state parameter $w=-1$. The
interaction between dark matter and neutrinos is given in terms of
the opacity $Q=\langle \sigma_{dm-\nu}|v| \rangle /m_{dm}$, the
ratio of the thermal averaged dark-matter--neutrino cross section
to the mass of the dark-matter particle.

The paper is organized as follows. In the next Section, we discuss
a class of models of neutrino--dark-matter interaction for both
scalar and spinor dark-matter candidates and obtain an estimate
for the opacity $Q$. In Section III, we outline the cosmological
consequences of a NIDM component, and we compare those predictions
with the latest data on galaxy clustering from the Sloan Digital
Sky Survey (SDSS). In Section IV, we consider astrophysical
constraints, particularly those from observation of neutrinos
from supernova 1987A. Finally, in the last Section, we report
our conclusions.

\section{The neutrino--dark-matter interaction}

The possibility of new neutrino interactions with exotic matter
fields and their cosmological implications have been recently
considered by many authors. Couplings with a light scalar or
pseudoscalar boson, as in the Majoron model
\cite{hannestad1,hannestad2,pierpaoli,dodelson}, can have sizeable
effects on the CMB and the power spectrum of large-scale structure
(LSS), and might lead to a neutrinoless Universe for a massless or
very light scalar field \cite{dodelson}.

A different class of interactions of neutrinos with DM, and more
generally of DM with the electromagnetic plasma as well, was put
forward in a series of papers discussing the possibility that DM
consists of particles with mass $m_{dm}$ in the MeV range
\cite{boehm1, boehm2, boehm3, hooper, boehm4}. If the relic
abundance of these particle is produced via the freezeout of
annihilation processes and yet corresponds to the observed DM
energy density today, they must interact with {\it stronger than
weak} interactions, since the cross section for annihilations via
the exchange of a massive particle as, e.g., a vector boson,
typically decreases at the freeze-out as the square of the
DM-particle mass.

The implications of this scenario for big bang nucleosynthesis
(BBN) have been considered in Ref.~\cite{serpico}.  With
DM-neutrino interactions producing the required DM annihilation
cross section of a few pico-barns (for an $s$-wave annihilation)
at decoupling, the mass range $m_{dm} \leq 10$ MeV is disfavored.
In fact, tightly-coupled DM particles give in this case a
non-negligible contribution to the total energy density, and the
neutrino-photon temperature ratio is increased because of the
entropy release from DM to neutrinos. Both effects conspire to
produce an order one extra effective neutrino species and thus an
increased primordial helium abundance, which is difficult to
reconcile with the present determination of the $^4$He mass
fraction; see, e.g., Refs.~\cite{cyburt1, cuoco, cyburt2,
serpico2}.

Neutrino coupling to DM particles with mass in the MeV range can
also be bound using neutrino fluxes from type II supernovae
\cite{sigl}, such as SN1987A, again resulting in a constraint
$m_{dm} \geq 10$ MeV. It is worth observing however, that this
bound only applies if these particles have large elastic
scattering cross sections with nucleons, larger than
neutrino-nucleon ones, whose effect is to shift outward the
neutrinosphere and thus leads to a lower neutrino decoupling
temperature.

In the following, we consider a DM-neutrino coupling even stronger
than those considered in Refs.~\cite{boehm1, boehm2, boehm3, hooper,
boehm4}, so that the resulting large scattering cross sections
might lead to detectable effects in the LSS power spectrum. In fact,
as already emphasized in Ref.~\cite{boehm2}, a DM-neutrino coupling
which leads to an order pico-barn annihilation cross section can
only affect the power spectrum on very small scales, since
scattering processes freeze out quite early in time. In this case,
LSS forms as in the presence of the usual collisionless DM fluid,
while neutrinos act via free streaming and suppress power on
scales smaller than the acoustic horizon at the time they become
non-relativistic.

We assume $m_{dm} \geq 10$ MeV, since our large DM-neutrino
coupling implies that the bound obtained in Ref.~\cite{serpico}
applies {\it a fortiori}. Notice that our working hypothesis
potentially leads to the problem of how a sizeable relic abundance
of DM particle actually forms, since a stronger coupling with
neutrinos enhances their annihilation rate and leads to negligible
relic abundance today. We will comment on this point later and
emphasize that our scenario requires an asymmetry between DM
particle and antiparticle produced at an early stage.

The interaction Lagrangian density depends upon the spin content
of the DM field, denoted in the following by $\psi$. We consider
here the case of a non-self-conjugated scalar particle or a Dirac
spinor and interaction terms that admit a conserved global $U(1)$
charge with $\psi$ transforming with a phase under the
corresponding transformations. For a scalar $\psi$ we can write
\be {\cal L}_{int}= h \overline{F}_R \nu_L \psi + h.c. \vv
\label{lscalar} \ee where $F$ is a spinor field. Similarly, DM
coupling to neutrinos can be also introduced via the interaction
with an intermediate vector-boson field $U_\mu$, \bea {\cal
L}_{int} &=& i g_\psi \left( \psi^*
\partial^\mu \psi - \psi \partial^\mu \psi^* \right) U_\mu +
g_\psi^2 \psi^* \psi U_\mu U^\mu \nonumber \\ &+& g_\nu
\overline{\nu}_L \gamma^\mu \nu_L U_\mu \pp \label{lscalaru} \eea
In both cases, we consider the $F$ or $U$ field to have masses of
order of MeV or larger to forbid $\psi$ decays at tree level,
which would erase any relic abundance of $\psi$ unless the
couplings are tuned to be very small. In a way, this assumption is
far from being {\it ad hoc} for the case of a coupling as in
Eq.~(\ref{lscalar}). If the $F$ field is lighter than $\psi$, we
simply have to shift our perspective and consider $F$ as a
candidate for light dark matter rather than the $\psi$ field.

In the range of neutrino temperature $T \leq MeV$ we are
interested in, the thermally averaged $\psi$-neutrino scattering
cross section is \be \langle \sigma_{dm-\nu} |v| \rangle \sim
|h|^4 \frac{T^2}{(m_F^2-m_{dm}^2)^2} \vv \label{sigmafs} \ee for
an $F$ fermion exchange, unless the $\psi$ and $F$ fields are
degenerate in mass, in which case the low-energy transfer
scattering cross section gets the usual Thomson behavior, \be
\sigma _{dm-\nu} \sim \frac{|h|^4}{m_{dm}^2}, \,\,\, m_F = m_{dm}
\pp \label{sigmafsdeg} \ee Similarly, for a $U$ coupling we have
\be \langle \sigma_{dm-\nu} |v| \rangle \sim g_\psi^2 g_\nu^2
\frac{T^2}{m_U^4} \pp \label{sigmaus} \ee The Lagrangian density,
Eq.~(\ref{lscalar}), also describes the interaction of (chiral)
fermionic dark matter with neutrinos via the exchange of a massive
scalar field, with the obvious redefinition $\psi \leftrightarrow
F$, and scattering cross sections are again given in this case by
Eqs.~(\ref{sigmafs}) or (\ref{sigmafsdeg}).

Finally, interaction of a Dirac DM field with neutrinos via a
vector-boson interaction Lagrangian, \bea {\cal L}_{int} &=& g_\psi
\left( c_L \overline{\psi}_L \gamma^\mu \psi_L +
c_R\overline{\psi}_R \gamma^\mu \psi_R \right) U_\mu \nonumber
\\ &+& g_\nu \overline{\nu}_L \gamma^\mu \nu_L U_\mu \vv
\label{lfermionu} \eea gives, for $m_{dm}\geq T$, \be \langle
\sigma _{dm-\nu} |v| \rangle  \sim g_\psi^2 g_\nu^2
\left(c_L^2+c_R^2-c_L c_R \right) \frac{T^2}{m_U^4} \pp
\label{sigmauf} \ee

A light $U$ boson with an order MeV
mass coupled to charged leptons might affect the electron-neutrino
scattering cross section at low energy, while
measurements shows no significant deviations from the standard
electroweak-model result \cite{allen,lsnd} (see also
Ref.~\cite{boehm2} for a detailed analysis on this issue). A possible
way out is of course to suppress the value of the $U$ coupling to
ordinary matter, including neutrinos, with respect to its coupling
to DM particles. In this case, however, neutrino--dark- matter
scattering would be quite small and again no observable effects on
LSS can be obtained. Another possibility, though less appealing,
is to assume that the $U$ boson couples mainly to neutrinos and very
weakly to charged leptons.

Regardless of the particular nature of DM particles and the
particular coupling to neutrinos, provided their mass as well as
the mass of the exchanged particle is in the range of MeV or
larger, we see from our discussion that the typical thermally
averaged scattering cross section with neutrinos for $T \leq$ MeV
has two possible distinct behaviors, either decreasing as $T^2$ or
constant for mass degeneracy of DM and intermediate scalar/fermion
particle $F$. In particular, it is useful to define the
DM-neutrino opacity, the thermally averaged scattering cross
section over DM mass ratio, as follows \be \frac{\langle
\sigma_{dm-\nu} |v| \rangle}{m_{dm}} \equiv Q_2 \frac{1}{a^2} \vv
\label{qu2} \ee for the $T^2$ behavior as in Eqs.~(\ref{sigmafs}),
(\ref{sigmaus}), or (\ref{sigmauf}). Here, $a$ denotes the scale
factor, normalized to unity at the present time.

The opacity can be written in terms of the DM-neutrino coupling,
which will be generically denoted by $g$ and a mass scale $M$ of
order MeV or larger. For example, for a scalar DM particle coupled
to neutrinos via the exchange of a fermion particle $F$ we have
$g=|h|$ and $M^2=|m_F^2-m_{dm}^2|$ [see Eq.~(\ref{sigmafs})],
while for a $U$ exchange, $g^2=g_\psi g_\nu$ and $M=m_U$ [see
Eq.~(\ref{sigmaus})]. In this notation, and using the known value
of neutrino temperature today, we get \be Q_2 \sim
\frac{g^4}{(M/{\rm MeV})^4} \frac{1}{m_{dm}/{\rm MeV}} \cdot
10^{-41} {\rm cm^2}~{\rm MeV}^{-1}. \label{q2bis} \ee

Similarly, for the case of a constant scattering cross section as
in Eq.~(\ref{sigmafsdeg}) we define \be \frac{\langle
\sigma_{dm-\nu} |v| \rangle}{m_{dm}} \equiv Q_0 \label{q0} \vv \ee
which gives \be Q_0 \sim \frac{|h|^4}{(m_{dm}/{\rm MeV})^3} \cdot
10^{-22} {\rm cm^2}~{\rm MeV}^{-1} \pp \label{q0bis} \ee Upper
bounds on these parameters will be discussed in the following
Section using the LSS power spectrum as well as other
astrophysical constraints.

We now come back to the issue of the relic abundance of DM
particles $\psi$. In the usual scenario, the present value of the
DM energy density results from the freezeout of annihilation
processes at temperatures of the order of $m_{dm}/20$; see, e.g.,
Ref.~\cite{kt}. As we will see in the next Section, the key
parameter entering the Euler equation ruling the DM fluid
perturbation is the effective DM-neutrino scattering rate, defined
by \be \Gamma_{sc} \sim  \frac{{\rho}_\nu}{{\rho}_{dm}} n_{dm}
\sigma_{dm-\nu} \vv \label{gammascattering} \ee with $n_{dm}$ the
DM number density and $\rho_\nu$ and $\rho_{dm}$ the energy
density of neutrinos and DM particles, respectively. If we assume
that $m_{dm}\geq 10$ MeV and that scattering processes with
neutrinos are still effective for small temperatures $T\ll$MeV,
thus leaving an imprint on the LSS power spectrum and CMB,
annihilation to neutrinos would reduce the DM energy density to a
very tiny value today.

To show this let us assume that indeed the annihilation processes
$\psi \overline{\psi} \rightarrow  \nu \overline{\nu}$ freeze out
at temperature of the order of $T_f=m_{dm}/20$\footnote{In
general, DM particles are coupled to the electromagnetic plasma as
well and annihilate into $e^+e^-$ pairs. In this case, the
corresponding annihilations will be assumed to freeze out at $T_f$
or earlier.}. We first consider the case of a
(non-self-conjugated) scalar DM field. For the coupling to
neutrinos via the exchange of a massive fermion $F$ [see
Eq.~(\ref{lscalar})], we get for the annihilation cross section,
\be \sigma_{ann} \sim |h|^4
\frac{m_{dm}^2}{(m_F^2+m_{dm}^2)^2}\frac{T}{m_{dm}} \pp \ee Notice
that the $s$-wave annihilation contribution is suppressed by the
square of neutrino mass, and has been neglected. This is a
consequence of the fact that the coupling is chiral; see, e.g.,
Ref.~\cite{xuelei}. Using Eq.~(\ref{sigmafs}), we see that at
$T_f$ the ratio of the thermally averaged neutrino scattering rate
over the annihilation rate $\Gamma_{ann} = \sigma_{ann} |v|
\,n_{dm}$ is of the order of \be \frac{\Gamma_{sc}}{\Gamma_{ann}}
\sim
\frac{\left(m_F^2+m_{dm}^2\right)^2}{\left(m_F^2-m_{dm}^2\right)^2}
\frac{T_f}{m_{dm}} \vv \ee when $m_F \neq m_{dm}$ [see
Eq.~(\ref{sigmafs})] or \be \frac{\Gamma_{sc}}{\Gamma_{ann}} \sim
4 \frac{m_{dm}}{T_f} \vv \ee when $m_F = m_{dm}$ [see
Eq.~(\ref{sigmafsdeg})]. This means that if annihilation freezes
out at $T\sim m_{dm}$, neutrino scatterings on DM particles are
also largely ineffective at temperature smaller than MeV. In this
case, no signatures of neutrino-DM interactions can be constrained
by present observations. The largest scale where DM-neutrino
interactions can leave an imprint corresponds to a wavenumber \be
k \sim \frac{2 \pi}{c} \frac{H(z_{sc})}{1+z_{sc}} \sim
\frac{1+z_{sc}}{\sqrt{1+z_{eq}}} \cdot 10^{-3}\, h \,
\mbox{Mpc}^{-1} \vv \label{zsc} \ee where $H(z_{sc})$ is the
Hubble parameter at the redshift $z_{sc}$ where DM-$\nu$
interactions freeze out and $z_{eq}$ is the redshift of
matter-radiation equality. For $z_{sc} \sim 10^8$, corresponding
to $T \sim 0.01$ MeV, we obtain $k \sim 10^3 h$ Mpc$^{-1}$.

Let us now consider the case of a $U$-exchange interaction. The
right order of magnitude  (10 pb) of the ($p$-wave) annihilation
cross section requires a light $U$ boson mass, of the order of
MeV, and coupling $h \sim 10^{-3}$. For $T\leq m_{dm}$, we get \be
\sigma_{ann} \sim g_\psi^2 g_\nu^2 \frac{m_{dm}^2}{(m_U^2-4
m_{dm}^2)^2} \frac{T}{m_{dm}} \pp \ee Comparing this result with
Eq.~(\ref{sigmaus}), we get at $T_f$, \be
\frac{\Gamma_{sc}}{\Gamma_{ann}} \sim
\frac{\left(m_U^2-4m_{dm}^2\right)^2}{m_U^4} \frac{T_f}{m_{dm}}
\vv \ee Again neutrino scatterings freeze out quite early around
$T_f$.

For spin-1/2 DM (Dirac or Majorana) particles and $F$ or $U$
couplings of Eqs.~(\ref{lscalar}) and (\ref{lfermionu}), one can
reason along the same lines, with similar results.

The case of a neutral scalar DM particle $\psi$ is quite
different. In this case in fact, the $F$ coupling to neutrinos
corresponds to an annihilation cross section which vanishes in the
limit of massless neutrinos. On the other hand, couplings of DM to
charged leptons can produce the correct relic abundance if $m_F$
is of the order of 100 GeV to 1 TeV \cite{boehm2}. However, for
such large values of $m_F$, the scattering cross section is very
small too (in particular it vanishes in the local limit $m_F
\rightarrow \infty$ due to a cancellation between the $s$- and
$u$-channel amplitudes), so the case of self-conjugated
scalar-DM--neutrino coupling is of no interest for the purposes of
this paper, and is not expected to produce any observable features
in the LSS.

Summarizing, if we assume that DM couples to neutrinos strongly
enough to produce observable effects that can be constrained by
CMB and LSS observations, we have to abandon the idea that relic
DM density formed via the usual mechanism based on freezing of DM
annihilation processes at temperatures $T \sim m_{dm}$.

How can our scenario be reconciled with the observed DM
contribution to the present energy density of the Universe? First
of all, it should be mentioned that only a fraction of the total
DM could be coupled to neutrinos. In fact, we will consider this
case too in the following. However, this would not represent a
solution to the problem, since this neutrino-coupled component
would completely annihilate into neutrinos at $T_f$. The more
interesting possibility is therefore that there is a
particle-antiparticle asymmetry produced at higher temperatures in
the DM sector coupled to neutrinos, very much like the mechanism
by which the baryon (and lepton, in the framework of leptogenesis)
number is produced in the early Universe. Indeed, this possibility
is also motivated by the intriguing observation that the
parameters $\Omega_b$ and $\Omega_{dm}$ only differ by a factor
five today, yet their production mechanism is usually considered
to be quite distinct, with very few exceptions \cite{kaplan,
Hooper:2004dc}; see also the discussion on this point in
Ref.~\cite{boehm2}. Though a more detailed theoretical analysis of
this possibility is perhaps still needed, we think that this idea
still represents a stimulating possible scenario. In this case, it
is meaningful to check to what extent present data can constrain
strong couplings of DM particles with mass in the range MeV or
higher to neutrinos. This scenario requires that DM particles are
excitations of a non-self-conjugated field, such as a complex
scalar field or a Dirac field, and that a particle-antiparticle
asymmetry in the DM sector
$\eta_\psi=(n_\psi-n_{\overline{\psi}})/n_\gamma$ has been
produced at some early stage, of the order of \be \eta_\psi =
\eta_B \frac{\Omega_{dm}}{\Omega_b} \frac{m_p}{m_{dm}} \vv \ee
with $m_p$ the proton mass and $\eta_B$ the baryon-to-photon
ratio, $\eta_B \sim 6.3 \times 10^{-10}$; see e.g.,
Ref.~\cite{serpico2}. The analysis performed in the following
Sections relies on these assumptions.

We conclude this Section by reporting the expression of the
smaller-wave mode $k$ for which we expect to see the effects of
DM-$\nu$ scatterings in the dark-matter perturbation in terms of
the opacities $Q_2$ and $Q_0$ introduced in Eqs.~(\ref{q2bis}) and
(\ref{q0bis}). This might be useful to understand the results
reported in the following. From the definition of $z_{sc}$, \be
\Gamma_{sc} (z_{sc}) \sim H(z_{sc}) \vv \ee and using standard
values for the neutrino temperature today and assuming no extra
relativistic degrees of freedom in addition to photons and
neutrinos, it is easy to get from Eq.~(\ref{zsc}) \be k \sim  0.2
\left(\frac{ 10^{-41} {\rm cm}^2~{\rm MeV}^{-1}}{Q_2}\right)^{1/4}
h \,{\rm Mpc}^{-1} \vv \label{kminq2} \ee or, for a constant
scattering cross section, \be k  \sim  0.2 \times 10^{-5} \left(
\frac{10^{-22} {\rm cm}^2~{\rm MeV}^{-1}}{Q_0} \right)^{1/2} h \,
{\rm Mpc}^{-1}. \label{kminq0} \ee

\section{NIDM and structure formation}

In order to include a neutrino--dark-matter interaction, we modify
the standard Euler equations resulting in (using conformal
time),
\begin{equation}
\label{cdm2} \dot{\theta}_{dm} = -{\dot{a}\over a}\theta_{dm} +
{4\rho_\nu \over 3\rho_{dm}}a n_{dm}\sigma_{dm-\nu}
(\theta_\nu-\theta_{dm}) \vv\
\end{equation}
\begin{equation}
\label{neutr} \dot{\theta}_\nu = k^2 \left(\frac{1}{4}\delta_\nu -
\sigma_\nu\right) + a n_{dm} \sigma_{dm-\nu} (\theta_{dm} -
\theta_\nu) \vv
\end{equation}
where momentum conservation in scattering processes has been
accounted for. With $\theta$ we denote the velocity perturbations,
the subscripts $``dm$'' and $``\nu$'' standing for
neutrino-interacting dark matter and neutrinos,
respectively. The quantity
$a n_{dm} \sigma_{dm-\nu}$ is the differential opacity and gives
the scattering rate of neutrinos by dark matter.

As in the case of the baryon-photon interaction, we neglect both the shear
term $k^{2}\sigma$ and the term $c_{dm}^{2}k^{2}\delta$, where
$c_{dm}$ is the sound speed of the dark-matter fluid.

As seen in the previous Section, the coupling between neutrinos
and dark matter can be parameterized through a cross section that
either decreases as $a^{-2}$, or takes a constant value. In these
two cases, the parameters determining the DM perturbations are the
opacities (cross section to DM-mass ratios) $Q_2$ and $Q_0$ defined in
Eqs.~(\ref{qu2}) and (\ref{q0}).

\begin{figure}[t]
\begin{center}
\includegraphics[width=7.5cm]{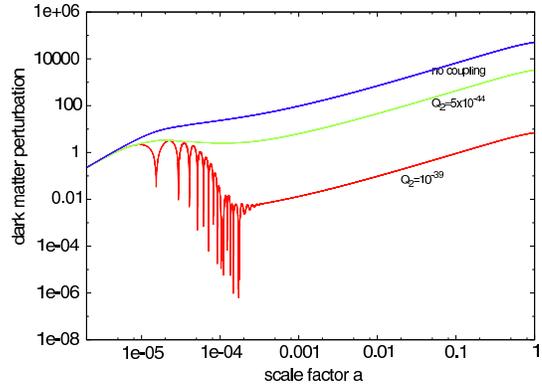}
\caption{Dark-matter perturbations of $k=1.04\,h\,$Mpc$^{-1}$;
the opacity $Q_{2}$ is in unit of cm$^{2}$~MeV$^{-1}$.  Damped oscillations are
clearly seen for $Q_2=10^{-39}$ cm$^{2}$~MeV$^{-1}$.} \label{fig:1}
\end{center}
\end{figure}

First consider the case $\langle \sigma_{dm-\nu}|v| \rangle
\propto a^{-2}$ (similar considerations can be made for the
constant cross section). In Fig.~\ref{fig:1}, we show what happens
when a perturbation of wavenumber $k=1.04\,h$ Mpc$^{-1}$ enters
the horizon for different values of $Q_2$. If the coupling is
zero, we have the standard picture. The mode enters the horizon in
the radiation-dominated era, and it starts to grow first
logarithmically and then linearly with the expansion factor
(during matter domination). When the same mode enters the horizon
with $Q_{2}=5\times\,10^{-44}$ cm$^{2}$~MeV$^{-1}$, the growth is
nearly zero during the radiation epoch, while the mode starts
growing linearly with the scale factor during matter domination,
since the coupling with neutrinos becomes negligible in this stage
for the chosen value of $Q_2$.

\begin{figure}[t]
\begin{center}
\includegraphics[width=7.5cm]{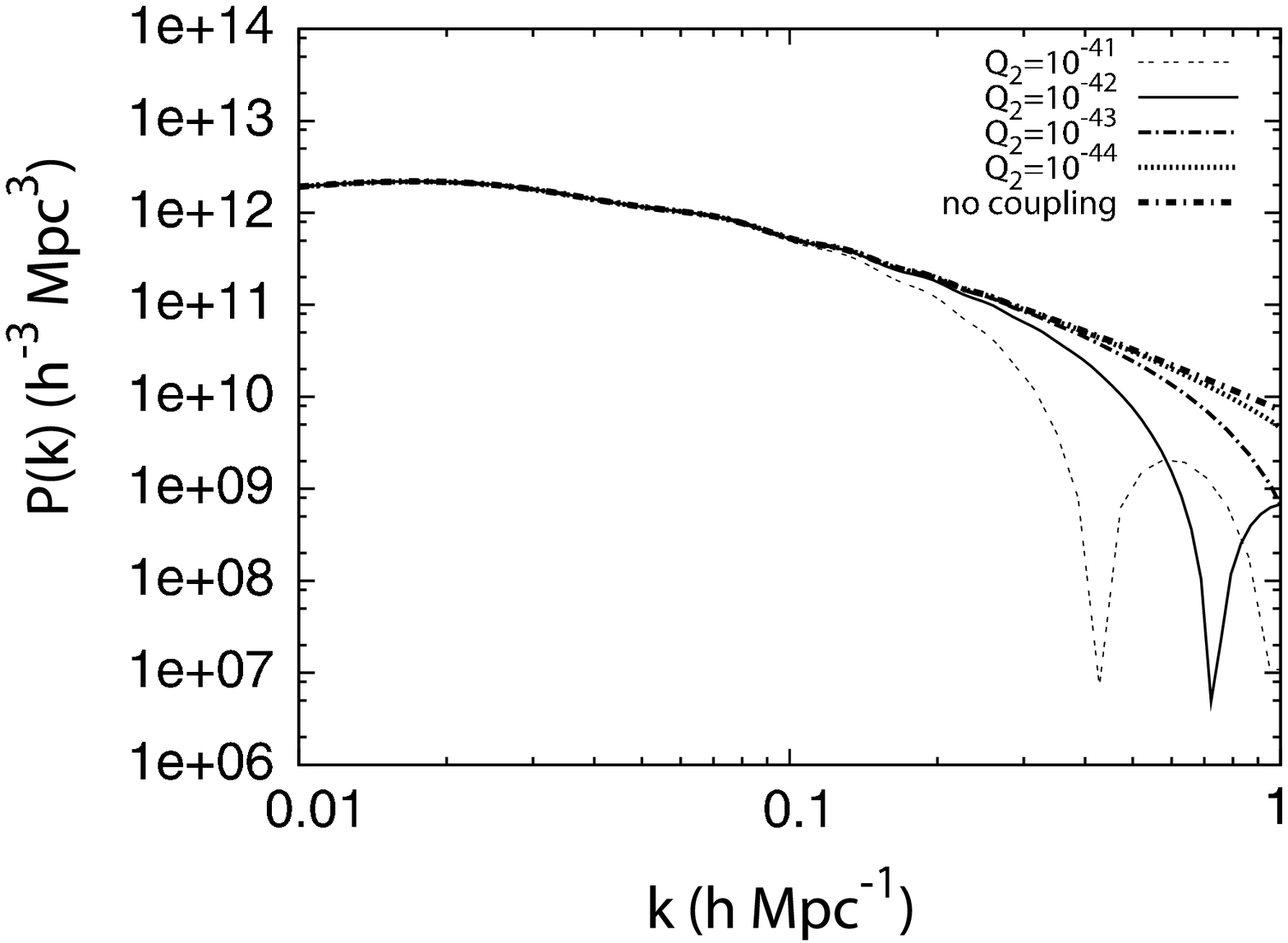}
\includegraphics[width=7.5cm]{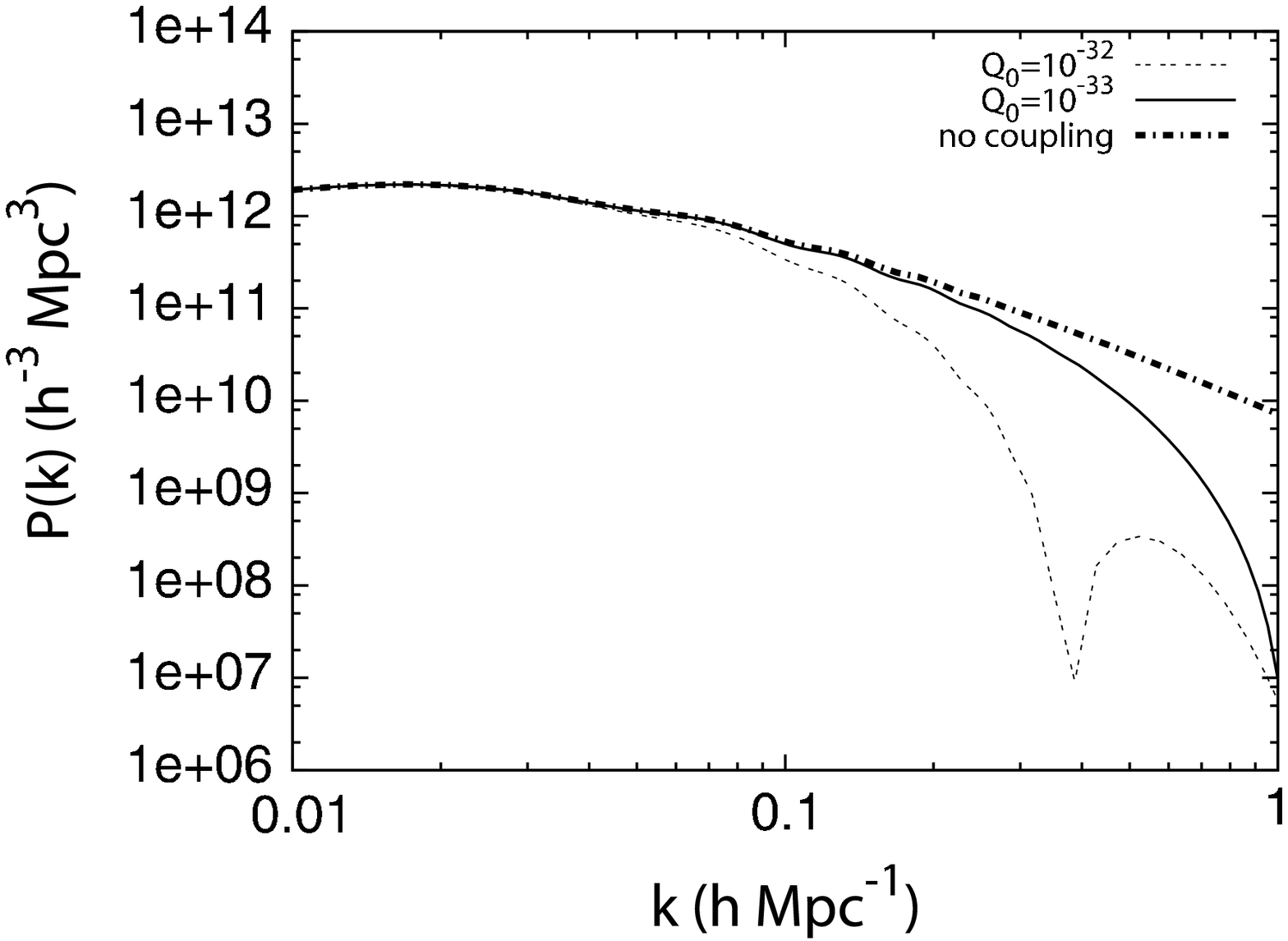}
\caption{Several matter power spectra with different opacities
$Q_{2}$ (top panel) and $Q_0$ (bottom panel) between dark matter
and neutrinos; $Q_{2}$ and $Q_0$ are in units of cm$^{2}$~MeV$^{-1}$.}
\label{fig:2}
\end{center}
\end{figure}

\begin{figure}[t]
\begin{center}
\includegraphics[width=7.0cm]{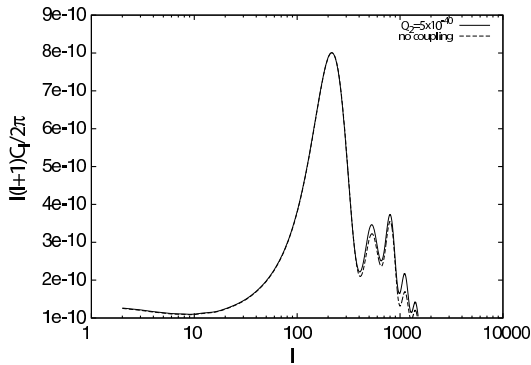}
\caption{Angular power spectra with and without
dark-matter--neutrino coupling. A small enhancement ($\sim
10\%$) of the height of the peaks on small scales is observed.} \label{fig:3}
\end{center}
\end{figure}

The situation is different when we consider a stronger coupling,
say $Q_{2}=10^{-39}$ cm$^{2}$~MeV$^{-1}$. When the perturbation enters the
horizon, dark matter is coupled with neutrinos and this results
in a series of oscillations until decoupling is reached.
Notice that the amplitude of oscillations decreases near
decoupling, because the decoupling itself is not instantaneous and
so we see diffusion damping for the dark-matter--neutrino fluid.
In Fig.~\ref{fig:2}, we plot several matter power spectra for
different values of the dark-matter--neutrino interaction for both
the couplings considered. The effect of the dark-matter--neutrino
interaction can be seen on small scales in the matter power
spectrum.  Larger couplings will correspond to later epochs of
neutrino-DM decoupling and to a damped oscillating regime on
larger scales. Finally, in Fig.~\ref{fig:3}, we plot the angular
power spectra of CMB anisotropies for two models with and without
DM-neutrino coupling.  For the value of $Q_2$ we
consider, which is already at odds with current clustering data,
there is a small enhancement in the small-scale CMB anisotropies.
The reason for this is that the anisotropic stress in the neutrino
relativistic component is reduced due to the coupling. In other
words, neutrinos are no more a fluid with a ``viscosity
parameter'' $c_{vis}^{2}=1/3$. This parameter,
introduced in Ref.~\cite{hu}, controls the relationship between
velocity/metric shear and anisotropic stresses in the neutrino
background \cite{melk}. The value of $c_{vis}^{2}$ will be close
to $c_{vis}^{2}=0$ and this implies a small enhancement of the
small-scale peaks, at the level of $~10\%$ \cite{melk}.
In order to bound the strength of DM coupling to neutrinos, we
consider the real-space power spectrum of galaxies in the Sloan
Digital Sky Survey (SDSS) using the data and window functions
of the analysis of Ref.~\cite{thx}. To compute the
likelihood function ${\cal L}^{SDSS}$ for the SDSS,
we restrict the analysis to a range of scales over which the
fluctuations are assumed to be in the linear regime ($k < 0.2
h^{-1}\rm Mpc$), and we marginalize over a bias $b$ considered to
be an additional free parameter. Since no relevant signature is
expected on CMB anisotropies for the values of $Q_2$ and
$Q_0$ we consider, we do not include CMB-anisotropy data in the
analysis and assume a cosmological concordance model with
$\Omega_{\Lambda}=0.70$ and $\Omega_{dm}=0.25$, which produces a
good fit to current CMB data.

By evaluating the SDSS likelihood we found that the couplings are
constrained to be
\bea Q_{2}&\le & 10^{-42}\,\mbox{cm}^2~\mbox{MeV}^{-1} \vv \label{resq2} \\
Q_{0}&\le& \,10^{-34} \,\mbox{cm}^{2}~\mbox{MeV}^{-1} \vv
\label{resq0} \eea at the $2 \sigma$ confidence level in above
fiducial cosmology. If we compare Eq.~(\ref{resq2}) with
Eq.~(\ref{q2bis}), we see that for couplings $g$ of order one this
bound is saturated if both $m_{dm}$ and $M$ are of the order of
MeV. Smaller values of $g$ imply lighter masses for the NIDM and
the intermediate exchanged particle in the scattering process. In
view of the BBN bound mentioned in Section II, $m_{dm} \geq 10$
MeV, these values are already disfavored, so the LSS constraint we
obtain is not further constraining the NIDM scenario. In fact, the
BBN can be weakened if we allows for more exotic features in the
neutrino density, in particular a neutrino chemical potential.
Indeed, in this case, the larger contribution of DM particles to
the Hubble expansion rate resulting into a higher $^4$He mass
fraction can be compensated by a positive (i.e., more neutrinos
than antineutrinos) value of the chemical potential;, see, e.g.,
Ref.~\cite{cuoco}. Further bounds might be obtained by studying
the LSS power spectrum at larger wavenumbers, $k \geq 0.2 \,h$
Mpc$^{-1}$, taking into account the nonlinear behavior of
perturbations for very small scales.

On the other hand, the result for $Q_0$ which we recall correspond
to a intermediate particle and NIDM mass degeneracy, is more
severely constraining $m_{dm}$. We get in this case $m_{dm} \geq
10 h^{4/3}$ GeV.

\section{Astrophysical Constraints}

It is interesting to compare Eqs.~(\ref{resq2}) and (\ref{resq0})
with the bounds on $Q_2$ and $Q_0$ that can be obtained from the
propagation of astrophysical neutrinos. The most important
constraint is provided by observation of neutrinos from SN1987A
\cite{raffelt}, which are in good agreement with the theoretical
expectation of neutrino fluxes from type II supernovae.  These
neutrinos have energies of order 10 MeV.  The thickness of the
dark-matter layer that they propagate through is approximately
$\int \rho(l) dl$, the integral of the dark-matter density along
the line of sight $l$ to the LMC.  Approximating the dark-matter
density $\rho(l) \sim \rho_0 (l/l_0)^{-2}$, where
$\rho_0\simeq0.4$ GeV~cm$^{-3}$ is the local density and
$l_0\simeq8$ kpc our distance from the Galactic center, we find a
dark-matter thickness $\sim10^{25}$ MeV~cm$^{-2}$.  Given the
agreement between the predicted and observed neutrino flux and
energy spectrum, we infer that neutrinos from SN1987A were not
significantly absorbed by dark matter along the line of sight,
from which we get an upper bound $\sim10^{-25}$ cm$^2$~MeV$^{-1}$
to the neutrino-DM opacity for neutrinos of energy $\sim10$ MeV.
From this result we obtain the upper bounds \bea Q_{2}&\le &
10^{-47}\left(\frac{10 {\rm MeV}}{M} \right)^2\,
\mbox{cm}^2~\mbox{MeV}^{-1} \vv
\label{resq2sn} \\
Q_{0}&\le& \,10^{-25}\, \mbox{cm}^{2}~\mbox{MeV}^{-1} \pp
\label{resq0sn} \eea We note that the bound on $Q_2$ is stronger
than what is obtained using LSS data and of the same order of
magnitude as the BBN limit corresponding to $M \geq 10$ MeV, while
for $Q_0$ the stronger bound is still provided by
Eq.~(\ref{resq0}).

Neutrinos with high energy are likely to be produced by a variety
of astrophysical sources. Strong scattering of these neutrinos off
the NIDM when traveling over cosmological distances of order of
tens of Mpc implies large energy losses and correspondingly a
strong deformation of the emitted energy spectrum at the source.
For light NIDM and intermediate exchanged particles (in the 10-MeV
range) the high-energy ($E_\nu \geq$ GeV) scattering cross section
behaves as $\sigma_{dm-\nu}(E_\nu \gg 10 {\rm MeV})\sim g^4/s$
with $s=m_{dm} E_\nu$. We stress once more that high values for
NIDM or intermediate $F$ or $U$ particle mass, though perfectly
legitimate, implies no observable effects in the LSS power
spectrum and are thus of no interest for the present analysis.
Using the definition of $Q_2$ and $Q_0$ of Section II and the
value of the critical density today, we can evaluate the typical
scattering length as a function of the neutrino energy as follows
\bea \frac{\lambda_\nu}{10 {\rm Mpc}} & \sim & \frac{E_\nu}{M}
\frac{1}{\Omega_{dm}} \frac{1}{Q_2} 10^{-44}\, {\rm cm^2~MeV^{-1}}, \\
\frac{\lambda_\nu}{10 {\rm Mpc}} & \sim & \frac{E_\nu}{M}
\frac{1}{\Omega_{dm}} \frac{1}{Q_0} 10^{-24}\, {\rm cm^2~MeV^{-1}}.
\eea If we use the LSS bounds of Eqs. (\ref{resq2}) and
(\ref{resq0}) we see that for $M \sim 10$ MeV, the value of
$\lambda_\nu$ is typically very large. The effect of interactions
with NIDM can only affect neutrinos with order-GeV energy over
distances of $10 \div 100$ Mpc assuming the largest value for
$Q_2$, while the effect is negligible for higher values of $E_\nu$
or for the $Q_0$ case.

\begin{figure}[t]
\begin{center}
\includegraphics[width=7.5cm]{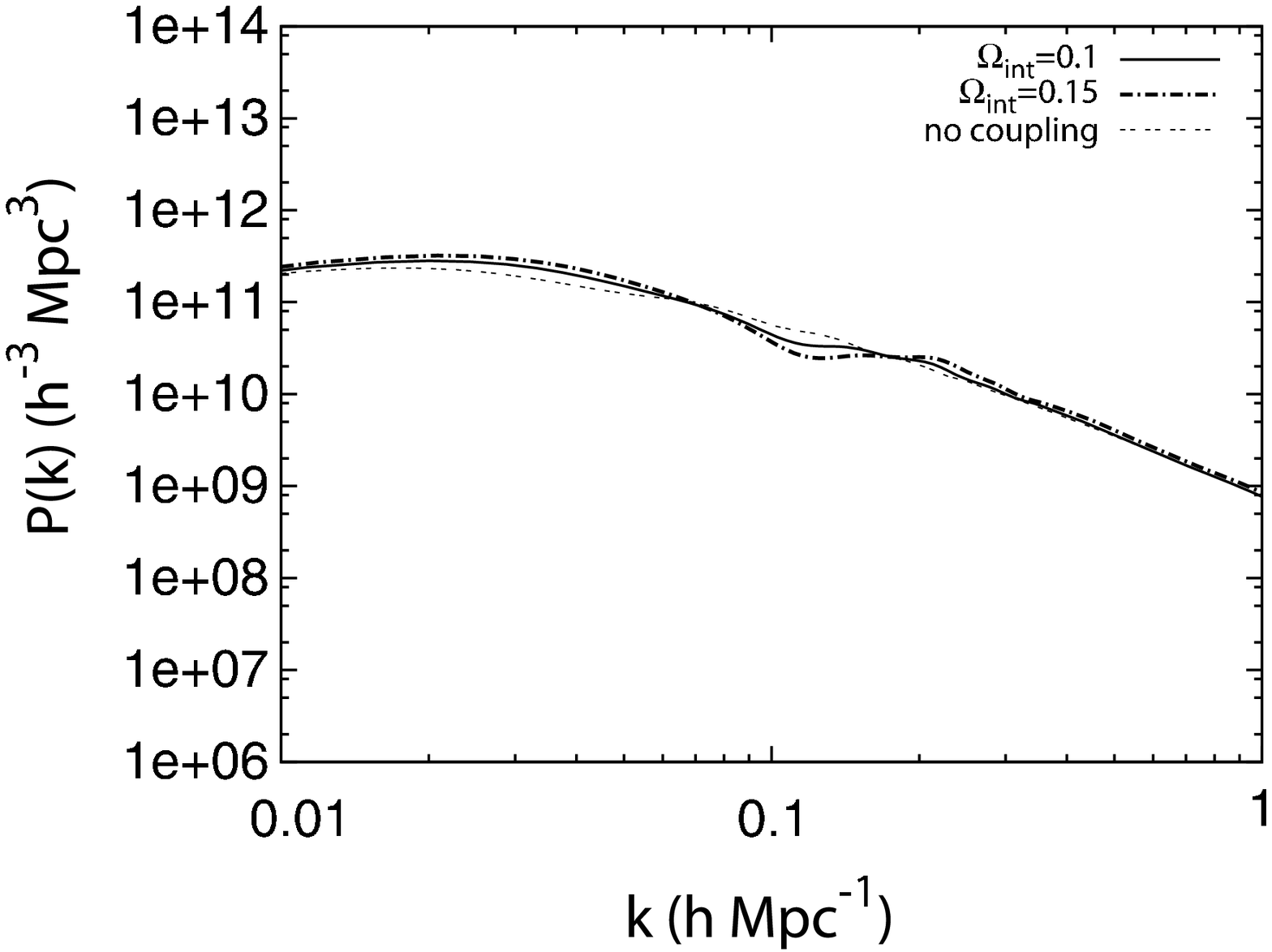}
\includegraphics[width=7.5cm]{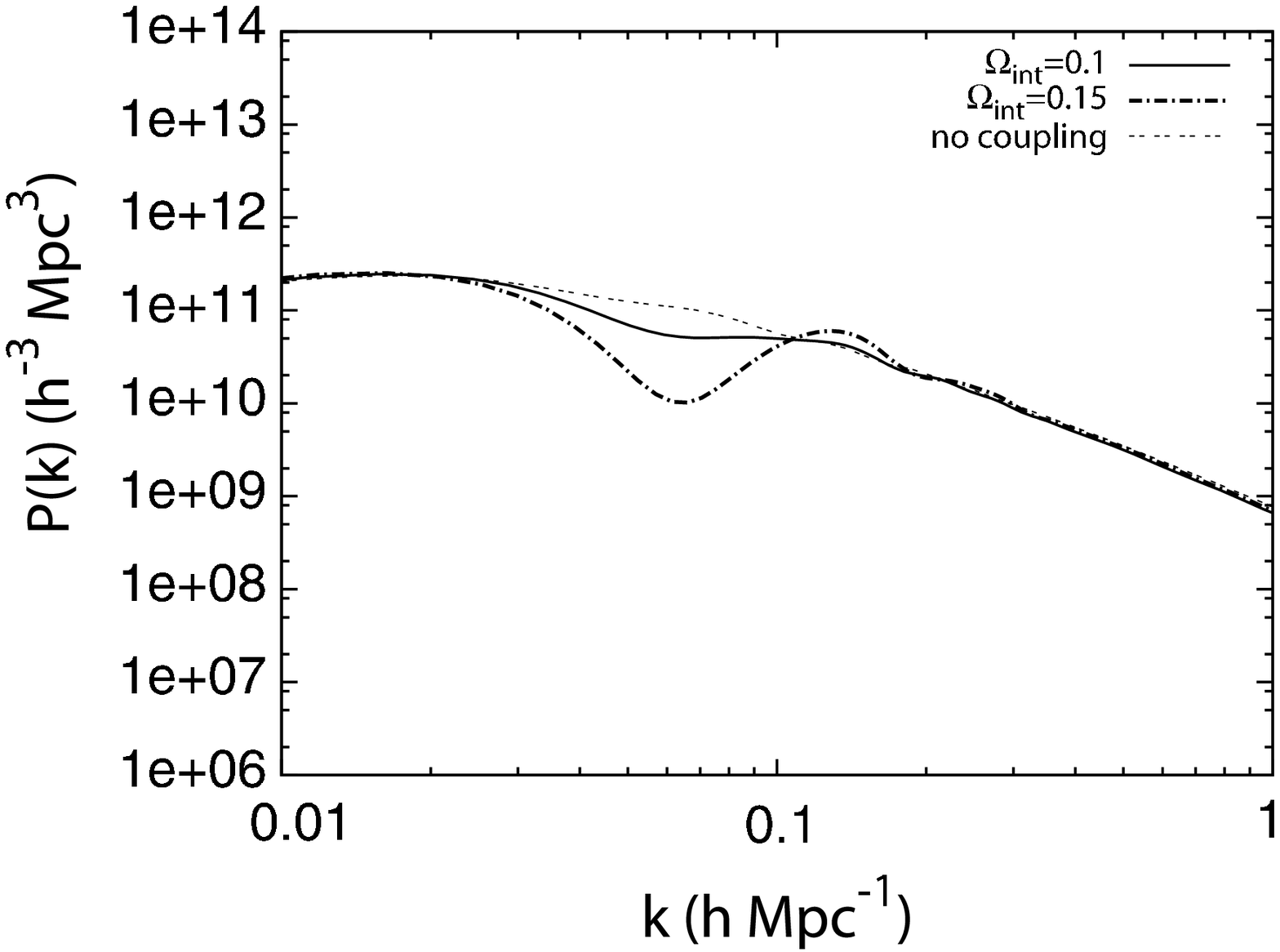}
\caption{Matter power spectra for cold+interacting dark matter
with $Q_{2}=10^{-38}$cm$^{2}$~MeV$^{-1}$ (top panel) and
$Q_2=10^{-37}$cm$^{2}$~MeV$^{-1}$ (bottom panel).} \label{fig:4}
\end{center}
\end{figure}

The bounds discussed so far are obtained under the assumption that
all dark matter is interacting with neutrinos. However, if the
dark matter is made of several components, it is possible that
only a fraction of the dark matter was actually strongly coupled.
In Fig.~\ref{fig:4}, we show matter power spectra for a standard
$\Lambda$CDM model with $\Omega_{dm}=0.25$ and
$\Omega_{\Lambda}=0.7$ and for other models where a fraction of
the energy density $\Omega_{int}= 0.7-\Omega_\Lambda$ of the
cosmological constant is replaced by interacting dark matter with
coupling $Q_{2}$. As we can see, the spectra are quite similar if
we consider quite large values of $Q_2$, as large as
$10^{-38}$cm$^2$~MeV$^{-1}$. This is simply due to the fact that
the interacting component is nearly unclustered on large scales,
$k \sim 0.01\,h$ Mpc$^{-1}$. Therefore, adding this component or
changing the energy density in $\Lambda$ is nearly equivalent.
This degeneracy in the framework of NIDM might weaken the current
estimates of the matter density from galaxy clustering. For higher
values of $Q_2$ or $\Omega_{int}$, oscillations in the power
spectrum are instead clearly visible.

\begin{figure}[t]
\begin{center}
\includegraphics[width=7.0cm]{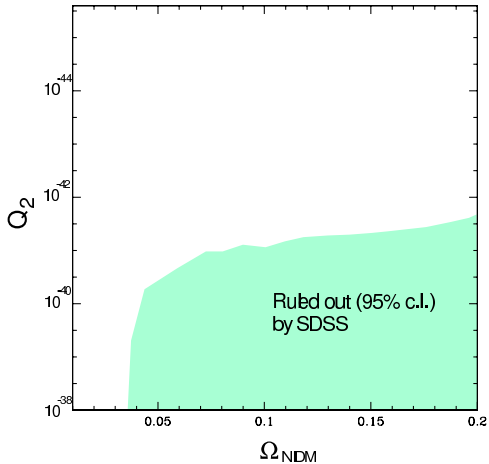}
\caption{Constraints on the $Q_2$ vs $\Omega_{NIDM}$ plane from
SDSS $P(k)$ measurements.
An overall matter density of $\Omega_m=0.27$ is assumed
with $\Omega_b=0.04$.} \label{fig:5}
\end{center}
\end{figure}

In Fig.~\ref{fig:5}, we plot constraints on $Q_2$ using SDSS $P(k)$ data
by allowing this possibility that only a fraction of the dark matter
interacts with neutrinos. The overall matter density is fixed at
$\Omega_m=0.27$ and we assume a flat universe. As we can see, a
smaller $\Omega_{NIDM}$ allows the possibility of relaxing the
constraints on $Q_2$.
%
\begin{figure}[t]
\begin{center}
\includegraphics[width=7.0cm]{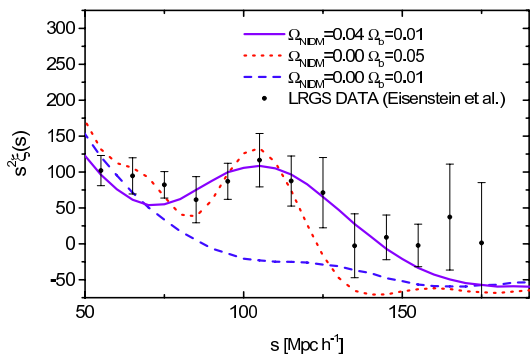}
\caption{Degeneracy for the baryon acoustic peak in the galaxy
correlation function. Three models are plotted: (1) a standard CDM
model with a baryon density that reproduces the baryon peak
(dotted line), (2) a standard CDM model with a low baryon density
(dashed line), and (3) a NIDM model with a low baryon density
(solid line). While a CDM model with a low baryon density fails to
describe the peak, a NIDM model with a low baryon density
describes the  peak adequately. The data are taken from the LRG
analysis of Eisenstein et al. 2005.} \label{fig:6}
\end{center}
\end{figure}
Recently, experimental evidence for a ``peak'' in the correlation
function of the SDSS luminous red galaxy (LRG) distribution at 100
Mpc scales has been reported \cite{eisen}. This peak is generally
interpreted as the imprint of oscillations in the photon-baryon
fluid near recombination. In the standard CDM framework, this peak
serves as an absolute ruler and with clustering measurements as a
function of redshift, one can extract strong constraints on the
dark-energy time evolution. However, it is clear that if a portion
of the dark matter is oscillating as a consequence of strong
coupling with neutrinos, that component may also produce an
oscillatory behavior and affect the conclusions of
Ref.~\cite{eisen}. In order to show this effect more
quantitatively we compare the LRG data  in Fig.~\ref{fig:6} with
the correlation function computed under three model descriptions:
a standard CDM model with a baryon density that reproduces the
baryon peak, a combination of a standard CDM model with a low
baryon density, and a model with a NIDM component and a low baryon
density\footnote{As suggested in Ref.~\cite{eisen}, using N-body
simulations, to compare between measurements in the redshift space
and predictions in the real space, we multiply the predicted
correlation functions by a conversion factor of
$[1+0.06/(1+(0.06s)^6)]^2$, where $s$ is the physical scale
measured in $h^{-1}$ Mpc}.

As we can see, NIDM is able to mimic the baryonic peak even in the
case of a low baryon density universe though in the standard CDM,
a low baryon density will not describe the oscillation. While the
possibility shown in Fig.~\ref{fig:6} is certainly fine tuned, one
may need to allow for such a scenario when deriving constraints on
the baryon energy density $\Omega_bh^2$ from large-scale-structure
observations if departures from the standard description were to
be considered. Let us note, however, that the above scenario,
while mimicking baryonic oscillations in galaxy clustering with
NIDM, would lead to a different shape for the CMB anisotropy power
spectrum. A combined analysis is therefore a powerful tool for
detecting NIDM in mixed models and any discrepancy between the
value of the baryon density derived independently from those
datasets could hint for an NIDM component.

\section{Discussion and Conclusions}

In this paper, we have studied the cosmological consequences of a
possible coupling between neutrinos and light dark matter with
mass in the MeV range. We considered two possible behaviors for
the thermally-averaged neutrino-DM elastic-scattering cross
section, either decreasing with temperature as $T^2$ or constant.
We compared the NIDM scenario with the large-scale galaxy
distribution and obtained upper limits on the opacity (ratio of
the DM-neutrino cross section to the dark-matter mass) of $Q_2 <
10^{-42}\, {\rm cm}^2$~MeV$^{-1}$ and $Q_0 < 10^{-34} {\rm
cm}^2$~MeV$^{-1}$ at the $95 \%$ C.L.
These limits may be relaxed if one consider the possibility that
only a fraction of the dark matter is made of NIDM. The main
cosmological observable for NIDM consists in diffusion-damped
oscillations in the matter power spectrum. Those NIDM oscillations
may affect current cosmological constraints on neutrinos masses
and dark energy from galaxy clustering. We have stressed that
strongly-coupled DM particles would have a non-negligible relic
abundance today only if an asymmetry between DM particle and
antiparticle is produced at some early stage in the evolution of
the Universe, since their density would vanish today because of
effective annihilation processes into neutrinos down to
temperatures much smaller than the DM mass. Detection of
NIDM-induced oscillations in the LSS power spectrum would be a
hint for such a non-standard scenario.

\begin{acknowledgments}
We acknowledge the use of the publicly available numerical code
\texttt{CMBFAST}. GM is pleased to thank P.~D. Serpico for
discussions. AM is supported by MURST through COFIN contract
no.~2004027755.  MK is supported by DoE DE-FG03-92-ER40701, NASA
NNG05GF69G, and the Gordon and Betty Moore Foundation.

\end{acknowledgments}

\end{document}